\documentclass[prl,twocolumn,showpacs]{revtex4}

\usepackage{graphicx}
\usepackage{amssymb}
\usepackage{amsmath}
\usepackage{color}
\usepackage[colorlinks]{hyperref}

\begin{document}

\title{Homeostatic Fluctuations of a Tissue Surface}

\author{Thomas Risler,$^{1,2}$ Aur\'elien Peilloux,$^{1,2}$ and Jacques Prost$^{1,2,3}$}
\affiliation{$^1$Laboratoire Physico Chimie Curie, Institut Curie, PSL Research University, CNRS, 26 rue d'Ulm, 75005 Paris, France}
\affiliation{$^2$Sorbonne Universit\'es, UPMC Univ Paris 06, CNRS, Laboratoire Physico Chimie Curie, 75005 Paris, France}
\affiliation{$^3$Mechanobiology Institute, National University of Singapore, 5A Engineering Drive 1, 117411 Singapore}

\newpage

\begin{abstract}
We study the surface fluctuations of a tissue with a dynamics dictated by cell-rearrangement, cell-division, and cell-death processes. Surface fluctuations are calculated in the homeostatic state, where cell division and cell death equilibrate on average. The obtained fluctuation spectrum can be mapped onto several other spectra such as those characterizing incompressible fluids, compressible Maxwell elastomers, or permeable membranes in appropriate asymptotic regimes. Since cell division and cell death are out-of-equilibrium processes, detailed balance is broken, but a generalized fluctuation-response relation is satisfied in terms of appropriate observables. Our work is a first step toward the description of the out-of-equilibrium fluctuations of the surface of a thick epithelium and its dynamical response to external perturbations.
\end{abstract}

\pacs{87.10.Mn, 05.40.Ca, 87.10.Ca, 87.18.Tt}

\maketitle

A standard way by which surfaces and interfaces are brought out of equilibrium is the generation of a flux, either orthogonal or parallel to their average plane, or a combination thereof~\cite{cates2000soft}. A familiar example of a parallel flux is that of wind blowing on the surface of water or more generally of an interface between two immiscible fluids submitted to hydrodynamic shear~\cite{Thiebaud:2010aa}. In the celebrated Kardar-Parisi-Zhang model~\cite{Kardar:1986dz}, which has been vastly studied by both physicists and mathematicians~\cite{halpin1995kinetic,hairer2013KPZ}, the flux is orthogonal to the growing interface. A flux can also be generated by the surface itself. This situation is achieved by biological membranes, in which proteins force an ion flow from one side of the membrane to the other, the energy source being provided by the hydrolysis of adenosine 5'-triphosphate (ATP)~\cite{prost1996shape,Ramaswamy:2000db,lacoste2014Active}. A similar situation is obtained with a membrane catalyzing the polymerization of a biopolymer on one of its sides~\cite{Maitra:2014aa}. Biological membranes, however, can be brought out of equilibrium without the existence of a flux by conformational changes of some of their constituents, which break detailed balance~\cite{lacoste2014Active}. In the case of red blood cells, for instance, this is achieved by ATP-dependent attachment and detachment of proteins that link the cytoskeleton to the phospholipids~\cite{Betz:2009aa}.

In this Letter, we propose a new way of driving an interface out of equilibrium. We consider the steady-state fluctuations of the surface of a tissue layer maintained in its homeostatic state, where cell division and cell death equilibrate on average~\cite{Basan:2009vn}. We use generic equations derived in Ref.~\cite{Ranft:2010uq}, in which stochastic cell division and cell death interfere with elastic stresses to provide effective viscoelastic equations. We neglect the mechanical role of the intercellular fluid and treat the tissue as a one-component system, which is valid for characteristic thicknesses smaller than a centimeter and time scales larger than a few minutes~\cite{Ranft:2012aa,delarue2014stress}. The reciprocal coupling between mechanical stress and cell renewal in the bulk of the tissue drives the system out of equilibrium. Despite their out-of-equilibrium nature, which breaks detailed balance and the fluctuation-dissipation relation, stochastic cell division and cell death can be described as Markovian processes~\cite{Ranft:2010uq}. Within this framework, the tissue is, therefore, expected to satisfy a generalized form of a fluctuation-response relation, which could potentially be tested experimentally~\cite{Prost:2009aa}.

\begin{figure}[h]
\begin{center}
\scalebox{0.3}{
\includegraphics{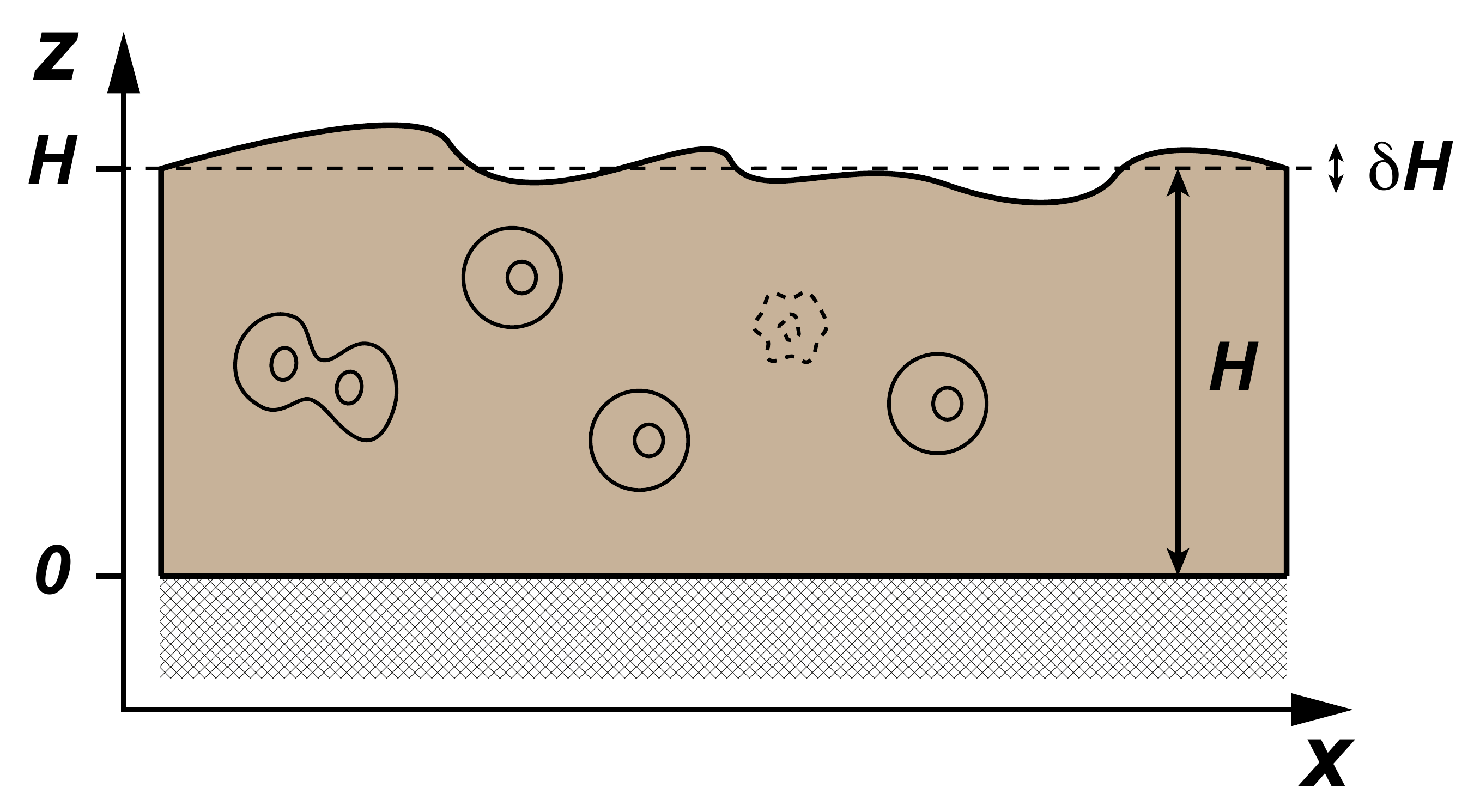}
}
\end{center}
\caption[fig1]
{\label{figGeometry} Illustration of the geometry of the system under study, with a schematic representation of the constitutive processes driving its out-of-equilibrium fluctuations. Cells in the tissue can be quiescent (smooth, rounded shapes), dividing (pinched, double-nucleated shape), or undergoing cell death (deformed and dotted cell and nucleus boundaries).}
\end{figure}

We consider an isotropic and homogeneous tissue of prescribed average thickness $H$ and infinite lateral extension resting on a rigid surface on one side and in contact with a classical fluid on the other (see Fig.~\ref{figGeometry}).
The fluid is of negligible viscosity and exerts a constant homogeneous pressure on the tissue, equal to the tissue homeostatic pressure~\cite{Basan:2009vn}. Under these conditions and on time scales shorter than the characteristic evolution time of the average thickness $H$ (see the Supplemental Material~\cite{SupplPRLrisl15}, Sec.~I), the tissue is in a steady state characterized by specific average values $\rho_h$ and $\sigma_h$ of the cell-number density $\rho$ and isotropic tissue stress $\sigma$ and by a vanishing cell-velocity field $v_\alpha$, where the index $\alpha$ stands for the spatial coordinates $x$, $y$, or $z$.

The cell-number density $\rho$ and the cell-velocity field $v_\alpha$ obey the continuity equation
\begin{equation}\label{eq:cont}
\partial_t\rho+\partial_\alpha(\rho v_\alpha)=(k_p+\xi_c)\rho \, ,
\end{equation}
where $\partial_t$ and $\partial_\alpha$ denote the partial derivatives with respect to time and spatial coordinates, with summation over repeated indices. In this equation, $k_p=k_d-k_a$ is the global mean rate of cell production, taking into account cell division and apoptosis with rates $k_d$ and $k_a$, and $\xi_c$ is a Gaussian white noise of zero mean. Cell division and cell death being independent stochastic processes, the variance of the noise reads $\langle\xi_c({\bf r},t)\xi_c({\bf r'},t')\rangle=[(k_d+k_a)/\rho]\delta({\bf r}-{\bf r}')\delta(t-t')$. The stress tensor $\sigma_{\alpha\beta}$ obeys the force-balance condition $\partial_\alpha\sigma_{\alpha\beta}=0$.

Close to the homeostatic state and to linear order, the deviation of the density from its homeostatic value $\delta\rho=\rho-\rho_h$ is proportional to $\delta\sigma=\sigma-\sigma_h$. The cell-production rate then reads $k_p=-(1/\tau_i)\,(\delta\rho/\rho_h)=(1/\tau_i)\,\chi_c\delta\sigma$, where $\tau_i$ is a characteristic time and $\chi_c$ is the short-time tissue compressibility in the homeostatic state. The stress tensor is decomposed into an isotropic and a traceless part $\sigma_{\alpha\beta}=\sigma\delta_{\alpha\beta}+\tilde{\sigma}_{\alpha\beta}$, where $\tilde{\sigma}_{\alpha\alpha}=0$. Under these conditions, the deviation $\delta\sigma_{\alpha\beta}=\delta\sigma\delta_{\alpha\beta}+\delta\tilde{\sigma}_{\alpha\beta}$ of the stress tensor with respect to its homeostatic value obeys the following Maxwell model~\cite{Ranft:2010uq}:
\begin{eqnarray}\label{eq:MaxwellModel}
\left(1+\tau_i\partial_t\right)\delta\sigma&=&\zeta \partial_\gamma v_\gamma-\xi  \nonumber\\
\left(1+\tau_a\partial_t\right)\delta\tilde{\sigma}_{\alpha\beta}&=&2\eta\,\tilde{v}_{\alpha\beta}-\tilde{\xi}_{\alpha\beta} \, .
\end{eqnarray}
In the isotropic part, $\zeta=\chi_c^{-1}\tau_i$ is an effective bulk viscosity, and $\xi=\zeta\xi_c$ is an effective bulk noise term of amplitude $\vartheta=\zeta^2(k_d+k_a)/\rho_h$. The remarkable absence of any compression modulus is due to the nonconservation of cell number~\cite{Basan:2009vn,Ranft:2010uq}. In the anisotropic part, $\tau_a$, $\eta$, and $\tilde{\xi}_{\alpha\beta}$ are the effective relaxation time constant, shear viscosity, and noise, {\it a priori} different from those appearing in the isotropic part, and $\tilde{v}_{\alpha\beta}=(1/2)\left[\partial_\alpha v_\beta+\partial_\beta v_\alpha-(2/3)\partial_\gamma v_\gamma\delta_{\alpha\beta}\right]$ is the traceless part of the velocity-gradient tensor. As done in Ref.~\cite{Ranft:2010uq}, we choose for $\tilde{\xi}_{\alpha\beta}$ a Gaussian white noise and write its second moment by symmetry arguments: $\langle\tilde\xi_{\alpha\beta}({\bf r},t)\tilde\xi_{\gamma\delta}({\bf r'},t')\rangle=\theta\left[\delta_{\alpha\gamma}\delta_{\beta\delta}+\delta_{\alpha\delta}\delta_{\beta\gamma}-(2/3)\delta_{\alpha\beta}\delta_{\gamma\delta}\right]\delta({\bf r}-{\bf r}')\delta(t-t')$, where $\theta$ characterizes the amplitude. In practice, this anisotropic noise is related to the stochasticity of cell rearrangements. In general, the two noises cannot be cast in a way respecting detailed balance with a common effective temperature, and the system is out of equilibrium. One can think of such a tissue as a system in which the particle number is not conserved and the isotropic and deviatory modes are maintained at different temperatures.

To characterize the position fluctuations $\delta H({\bf r},t)$ of the tissue upper surface, we focus our attention on the two-point autocorrelation function $C({\bf r}-{\bf r}',t-t')=\langle\delta H({\bf r},t)\delta H({\bf r'},t')\rangle$ and on its linear response function $\chi$ to an externally applied pressure field $\delta P_e({\bf r},t)$ defined as $\langle\delta H({\bf r},t)\rangle=\int \chi({\bf r}-{\bf r}',t-t')\delta P_e({\bf r}',t')\,d{\bf r}'dt'$. We further consider the Fourier components of the different fields defined according to the following Fourier transform: $f({\bf q},\omega)=\int d{\bf r}\,dt\,f({\bf r},t)\exp{[-i({\bf q}\cdot{\bf x}+\omega t)]}$. In a linear perturbation close to a flat interface, the different Fourier modes decouple, and we can consider the dependence on one single wave vector $q$ without loss of generality. Close to the homeostatic state, the equation governing the cell-velocity field reads in frequency space,
\begin{equation}
\eta_1\partial_\alpha\partial_\alpha v_\beta+\left(\zeta_1+\frac{1}{3}\eta_1\right)\partial_\beta(\partial_\alpha v_\alpha)=\partial_\alpha\xi_{1,\alpha\beta} \, ,
\end{equation}
with
\begin{equation}
\zeta_1=\frac{\zeta}{1+i\omega\tau_i} \, , \quad \eta_1=\frac{\eta}{1+i\omega\tau_a} \, ,
\end{equation}
and
\begin{equation}\label{eq:noise}
\xi_{1,\alpha\beta}=\frac{1}{1+i\omega\tau_i}\xi\delta_{\alpha\beta}+\frac{1}{1+i\omega\tau_a}\tilde{\xi}_{\alpha\beta} \, .
\end{equation}

Writing the stress continuity at the upper surface with generalized Laplace pressure and negligible shear stress, we obtain in Fourier space and in the linear regime:
\begin{align}\label{eq:BC}
&\left(\zeta_1-\frac{2}{3}\eta_1\right)\left(iq v_x+\partial_z v_z\right)+2\eta_1\partial_z v_z+(\gamma q^2+\kappa q^4)\,\delta H=\xi_{1,zz}\nonumber\\
&\textrm{and} \quad \eta_1\left(iq v_z+\partial_z v_x\right)=\xi_{1,xz}\, .
\end{align}
Here, $\gamma$ is the tissue surface tension and $\kappa$ its surface bending rigidity. To these must be added the kinematic condition $v_z=i\omega \delta H$. At its lower surface, the tissue is in contact with a hard wall, and we consider a no-slip boundary condition, such that $v_x=v_z=0$.

General expressions for the two-point autocorrelation function and the linear response function are given in the Supplemental Material~\cite{SupplPRLrisl15}, Sec.~II. Using these results, we characterize how far from thermodynamic equilibrium the system is operating by the introduction of an effective temperature $T_{\rm eff}$ equal to the thermodynamic temperature at equilibrium:
\begin{equation}\label{eq:effTempDef}
k_BT_{\rm eff}(q,\omega)=\frac{\omega}{2}\frac{C(q,\omega)}{\chi''(q,\omega)} \, .
\end{equation}
Here, $k_B$ is the Boltzmann constant, and $\chi''(q,\omega)$ is the imaginary part of the response function in Fourier space. The effective temperature is, in general, frequency and wavelength dependent, and its full expression is given in the Supplemental Material~\cite{SupplPRLrisl15}, Sec.~II. If we choose thermal noise amplitudes $\vartheta=2\zeta k_BT$ and $\theta=2\eta k_BT$, which define an equilibrium Maxwell model, the effective temperature equals the thermodynamic temperature $T$ at all frequencies. More generally, when $\vartheta/\theta=\zeta/\eta$, the effective temperature is constant and the system is analogous to an equilibrium system.

We first consider the limiting case where the thickness $H$ of the tissue is infinite or equivalently where $qH\gg 1$, for which full analytic expressions can be presented here. In this limit, the functions defined above read
\begin{eqnarray}\label{eq:CorrResp}
C(q,\omega)&\simeq&\frac{2}{q}\frac{\vartheta_1|\eta_1|^2+\theta_1\left(|\zeta_1+\frac{1}{3}\eta_1|^2+\frac{1}{3}|\eta_1|^2\right)}{\left|\zeta_1+\frac{4}{3}\eta_1\right|^2\,\left|\gamma q+\kappa q^3+2 i \omega \tilde\eta_1\right|^2}\nonumber\\
\chi(q,\omega)&\simeq&-\frac{1}{q}\frac{1}{\gamma q+\kappa q^3+2 i \omega \tilde\eta_1}\nonumber\\
k_BT_{\rm eff}(q,\omega)&\simeq&\frac{\vartheta_1|\eta_1|^2+\theta_1\left(\left|\zeta_1+\frac{1}{3}\eta_1\right|^2+\frac{1}{3}|\eta_1|^2\right)}{2\rm{Re}\left[\eta_1\left(\zeta_1+\frac{1}{3}\eta_1\right)\left(\zeta_1+\frac{4}{3}\eta_1\right)^*\right]} \, ,
\end{eqnarray}
with $\vartheta_1=\vartheta/[1+(\omega\tau_i)^2]$, $\theta_1=\theta/[1+(\omega\tau_a)^2]$ and $\tilde\eta_1=\eta_1(\zeta_1+\eta_1/3)/(\zeta_1+4\eta_1/3)$. The star denotes the complex conjugate. Note that, due to the absence of an intrinsic length scale in the system in this case, the effective temperature has no dependence on the wave vector. 

In the low-frequency limit ($\omega\tau_a,\omega\tau_i\ll 1$), these functions correspond to those of a passive compressible fluid with bulk and shear viscosities $\zeta$ and $\eta$ at the temperature $T_{\rm eff}$ but with the originality of a vanishing compression modulus. As already stated, the long-time compression modulus vanishes in a tissue due to the nonconservation of cell number~\cite{Basan:2009vn,Ranft:2010uq}, a feature that is impossible in a system with a conserved number of particles. Interestingly, the correspondence also holds with an incompressible fluid with effective shear viscosity $\tilde\eta=\eta(\zeta+\eta/3)/(\zeta+4\eta/3)$, since then compression modes are infinitely fast, and only viscous terms are left in the correlation and response functions. In the opposite limit ($\omega\tau_a,\omega\tau_i\gg 1$), the expressions are analogous to those of a passive, one-component, compressible Maxwell elastomer.

\begin{figure}[h]
\scalebox{0.35}{
\includegraphics{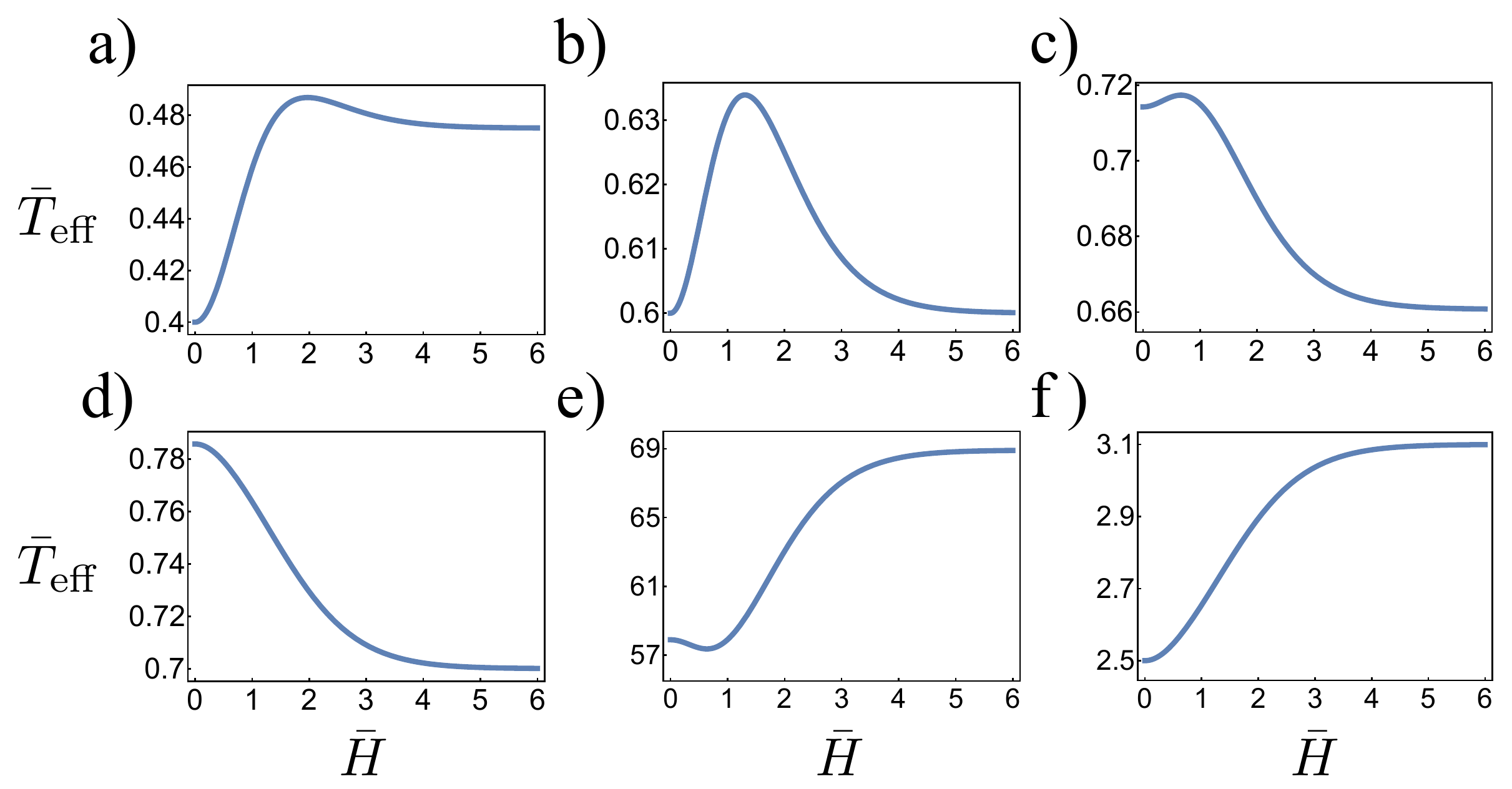}
}
\caption[fig2]
{\label{fig_effTemp} Wave-vector dependence of the effective temperature for six different parameter sets. In all six panels, we plot the adimensional ratio $\bar{T}_{\rm eff}=2\zeta k_BT_{\rm eff}/\vartheta$ as a function of $\bar{H}=qH$, for different values of $\vartheta/\theta$ and $\zeta/\eta$. In panels (a)--(d), $\vartheta/\theta-\zeta/\eta=1$, and $\zeta/\eta$ takes the values 1/3, 2/3, 1, and 4/3 in panels (a), (b), (c), and (d), respectively. In panels (e) and (f), $\vartheta/\theta-\zeta/\eta=-1$, and $\zeta/\eta$ equals 1.01 and 4/3, respectively. In panel (b), the asymptotic values in the low- and large-$q$ limits are both equal to 3/5.}
\end{figure}

In the case where the thickness $H$ of the tissue layer is finite, as stated above, the effective temperature is both frequency and wavelength dependent. In the low-frequency regime, it depends on wave vector only (see the Supplemental Material~\cite{SupplPRLrisl15}, Sec.~III for its expression). When $\vartheta/\theta\neq\zeta/\eta$, it can exhibit different nontrivial behaviors depending on the parameter values (see Fig.~\ref{fig_effTemp}).

Despite the existence of a frequency- and wave-vector-dependent effective temperature, the system's dynamics is Markovian. Proper observables, therefore, exist in terms of which a generalized fluctuation-response relation is satisfied~\cite{Prost:2009aa}. We choose here to illustrate this fact in the low-frequency domain ($\omega\tau_a,\omega\tau_i\ll 1$), where we can write a single evolution equation for $\delta H$, directly under a Markovian form. This allows us to present the explicit construction of an observable $X(q,t)$ that obeys a fluctuation-response relation. Solving for the integration constants and using the kinematic condition at the upper surface $v_z=\partial_t \delta H$, we get
\begin{equation}
\partial_t \delta H(q,t)=-\tau(q)^{-1}\delta H(q,t)+f(q,t)+\xi_{\delta H}(q,t) \, ,
\end{equation}
where $\tau(q)$ is a relaxation time constant that depends on the wave vector, $f(q,t)$ is an external field acting on the tissue upper surface, and the noise $\xi_{\delta H}(q,t)$ is a scalar combination of integrals containing the matricial noise $\xi_{\alpha\beta}=\xi\delta_{\alpha\beta}+\tilde{\xi}_{\alpha\beta}$ (see the Supplemental Material~\cite{SupplPRLrisl15}, Sec.~IV for explicit expressions). Introducing the equal-time autocorrelation function $\Sigma(q)$ of $\delta H(q,t)$ in the absence of an external field as $\langle \delta H(q,t)\delta H(q',t)\rangle=\Sigma(q)\delta(q+q')$, we define $X(q,t)$ as
\begin{equation}
X(q,t)=-\tau(q)\Sigma(q)^{-1}\,\delta H(q,t)  \, .
\end{equation}
With these definitions, we have~\cite{Prost:2009aa}
\begin{equation}
\chi_{XX}(q,\omega)-\chi_{XX}(q,-\omega)=i\omega C_{XX}(q,\omega) \, ,
\end{equation}
where $C_{XX}$ and $\chi_{XX}$ are, respectively, the two-point autocorrelation function of the variable $X$ and its linear response to the external field $f$, here written in Fourier space. The corresponding explicit expressions are given in the Supplemental Material~\cite{SupplPRLrisl15}, Sec.~IV.

Within the low-frequency domain, the regime $qH\ll 1$ is particularly interesting, since the expressions of the correlation and response functions are then analogous to those of a permeable membrane with permeation constant $\lambda_p=H/(\zeta+\frac{4}{3}\eta)$~\cite{prost1998fluctuation}:
\begin{eqnarray}\label{eq:membraneqomega}
C(q,\omega)&\simeq&\frac{1}{H}\frac{\vartheta+\frac{4}{3}\theta}{\left(\gamma q^2+\kappa q^4\right)^2+\omega^2\lambda_p^{-2}}\nonumber\\
\nonumber\\
\chi(q,\omega)&\simeq&-\frac{1}{\gamma q^2+\kappa q^4+i\omega\lambda_p^{-1}}\nonumber\\
\nonumber\\
k_BT_{\rm eff}&\simeq&\frac{1}{2}\frac{\vartheta+\frac{4}{3}\theta}{\zeta+\frac{4}{3}\eta} \, .
\end{eqnarray}
It is remarkable that, due to the nonconservation of cell number and the subsequent absence of long-time compression modulus, there is no lubrication regime. The leading term in the equal-time correlation function reads
\begin{equation}
C(q,t-t'=0)\simeq\frac{k_BT_{\rm eff}}{\gamma q^2+\kappa q^4}\, ,
\end{equation}
and we can make direct use of the results concerning the fluctuations of equilibrium membranes~\cite{lipowsky1995structure,nelson2004statistical}.

The divergence of the mean-square fluctuations of the tissue thickness is of particular interest. There are three length scales entering the problem. The first one is simply the thickness $H$. The second one is given by the square root of the ratio of the bending modulus over the membrane tension $l_\kappa=\sqrt{\kappa/\gamma}$. Finally, the third one, the ``collision'' length $l_c$, stems from the divergence of the mean-square fluctuations of the tissue thickness as a function of lateral position~\cite{fisher1982wall,helfrich1984undulations}. It is such that $\langle\delta H(0,t)\delta H(l_c,t)\rangle= H^2$. If $\gamma H^2\ll k_BT_{\rm eff}$, $l_c$ is proportional to the thickness $H$ and is much smaller than $l_\kappa$:
\begin{equation}
l_c\propto H \sqrt{\frac{\kappa}{k_BT_{\rm eff}}}=l_\kappa\sqrt{\frac{\gamma H^2}{k_BT_{\rm eff}}}\, .
\end{equation}
In this regime, the tissue surface fluctuations scale like distances parallel to the interface. In the tension-dominated regime where $\gamma H^2\gg k_BT_{\rm eff}$, $l_c$ reads
\begin{equation}
l_c\propto l_\kappa \exp\left(\frac{\pi\gamma H^2}{k_BT_{\rm eff}}\right) \, .
\end{equation}
In this regime, fluctuation amplitudes scale logarithmically with the distance parallel to the interface, and $l_c$ is much larger than $l_\kappa$. The surface appears locally smooth, although thickness variations eventually reach values of the order of the thickness itself on length scales of order $l_c$.

A natural question concerns the possibility to observe such effects. The effective temperature in the low-frequency, small-wave-vector regime can be estimated using Eq.~(\ref{eq:membraneqomega}). Keeping only the isotropic contribution in the homeostatic state, we obtain $k_BT_{\rm eff}=\zeta k_d/\rho_h$. Using this expression and numerical values from the literature, we estimate that the ratio $l_c/l_\kappa$ is typically of order one tenth to several tens and can potentially be much larger (see the Supplemental Material~\cite{SupplPRLrisl15}, Sec.~V). This indicates that the two limiting regimes mentioned here could, in principle, be observed, and that changing the physical properties of the tissue could affect its  surface structure without necessarily requiring a modification of the cell-division rate. The surface of precancerous tissues is usually rougher than that of healthy tissues, even when these tissues are not in a growth phase~\cite{weinberg2007bc,tavassoli2003pag}. One could speculate that some of the early cancerous mutations affect certain physical properties of the tissue, not necessarily uniquely related to cell proliferation.

In equilibrium membranes, the existence of a collision length generates an entropic repulsive interaction. In tissues, the physical meaning is likely to be very different. In the case we discuss here, this suggests that the tissue thickness potentially vanishes. The onset of tissue disappearance is given by the characteristic time of thickness fluctuations at the scale $l_c$:
\begin{equation}
\tau_d=\frac{\left(\zeta+\frac{4}{3}\eta\right)l_c^2}{\left(\gamma+\kappa l_c^{-2}\right)H}\, .
\end{equation}
The fate of the tissue layer then depends on the future evolution of the region of zero thickness. This, in turns, depends on the tissue-substrate interaction and adhesion properties, and a detailed analysis of this outcome is beyond the scope of this work. Note that the linear approximation used in this Letter is valid as long as $\kappa\gg k_BT_{\rm eff}$. In the opposite case, the surface would appear crumpled with no angular coherence~\cite{nelson2004statistical}.

The tissue surface fluctuations we have studied here exhibit a surprisingly rich behavioral diversity. One can define a wave-vector- and frequency-dependent effective temperature, a clear signature of the out-of-equilibrium nature of the system. In the low-frequency regime, this effective temperature is only wave-vector dependent. As shown in Fig.~\ref{fig_effTemp}, it can present different nontrivial behaviors, among which being monotonically varying or having an extremum at a finite wave vector. In the large-wave-vector, high-frequency regime, the tissue behaves like a compressible Maxwell elastomer with nonthermal noise, whereas in the large-wave-vector, low-frequency regime, it behaves like an incompressible fluid at equilibrium. Finally, in the small-wave-vector, low-frequency regime, the mapping can be made to a fully permeable membrane with tension and curvature. This implies that, at sufficiently large scales, the tissue always has thickness fluctuations on the order of the average thickness. This could signal a finite lifetime for the tissue layer. In actual epithelial tissues, however, such an outcome is usually not observed. In this case, cell division and cell death are regulated by external chemical gradients. As a consequence, cells are globally renewed close to the basal membrane and die close to the apical surface, which leads to a well-defined thickness~\cite{young2006wheater,Basan:2011fk,Risler:2013aa}. One could speculate that this is part of the solutions developed under evolutionary pressure to create robust epithelial tissues.


\newpage

\begin{widetext}

\begin{large}

{\noindent \bf SUPPLEMENTAL MATERIAL}\\

\bigskip

{\noindent \Large \bf Homeostatic Fluctuations of a Tissue Surface}\\

\bigskip

\noindent
Thomas Risler,$^{1,2}$ Aur\'elien Peilloux,$^{1,2}$ and Jacques Prost$^{1,2,3}$

\bigskip

\noindent
{\normalsize $^1$Laboratoire Physico Chimie Curie, Institut Curie, PSL Research University, CNRS, 26 rue d'Ulm, 75005 Paris, France}\\
{\normalsize $^2$Sorbonne Universit\'es, UPMC Univ Paris 06, CNRS, Laboratoire Physico Chimie Curie, 75005 Paris, France}\\
{\normalsize $^3$Mechanobiology Institute, National University of Singapore, 5A Engineering Drive 1, 117411 Singapore}

\section{I. Slow time evolution of the average thickness $H$}

In our study, we treat the average thickness $H$ as a constant in time. In order to justify this framework, we estimate here the characteristic time over which the average tissue thickness $H$ evolves by diffusion, due to stochastic cell-divison and cell-death processes, with average rates $k_d=k_a$. We consider a tissue layer made of $N$ cells of average cell-number density $\rho$, charateristic lateral dimension $L$ and average thickness $H$. In this volume, the cell-number standard deviation in a time $dt$ is $\delta N=\sqrt{(k_d+k_a)\,N\,dt}$. For a total volume $V=H\,L^2=N/\rho$, the corresponding amplitude of thickness fluctuations reads:
\begin{equation}
\delta H^2=\frac{(k_d+k_a)\,H}{\rho\,L^2}\,dt \, .
\end{equation}
To diffuse over distances of order $H$, the tissue layer therefore takes a characteristic time $\tau_H \simeq \rho\,H\,L^2/(k_d+k_a)$. With a maximal rate of cell renewal of one per day, a cell-diameter of 10 $\mu$m, a tissue thickness of 100 $\mu$m, this characteristic time of the order of several hundred years for a tissue patch of one millimeter square, and several tens of thousand years for one of one centimeter square. Ignoring the temporal variations of the average thickness $H$ is therefore perfectly valid in the framework of our study.

\section{II. Expressions of the physical quantities for the tissue of finite thickness $H$}

The two-point autocorrelation function reads
\begin{equation}\label{corrComplex}
C(q,\omega)=\frac{2}{q}\frac{\vartheta_1N_{a,1}+\theta_1N_{b,1}}{\left| (\gamma q +\kappa q^3)D_{a,1}+2i\omega\eta_1D_{b,1}\right|^2} \, ,
\end{equation}
where
\begin{eqnarray}\label{corrQuantitesComplex}
N_{a,1}&=&\left|2+\hat\zeta_1\right|^2\left(e^{8\bar H}-1\right)+4\bar H|\hat\zeta_1|^2e^{2\bar H}\left(e^{4\bar H}+1\right)+8\bar H\left[2+2(\hat\zeta_1+\hat\zeta_1^*)+(1-2\bar H^2)|\hat\zeta_1|^2\right]e^{4\bar H}\nonumber\\
&&+2\left(-2+|\hat\zeta_1|^2+2\bar H^2[2(\hat\zeta_1+\hat\zeta_1^*)+3|\hat\zeta_1|^2]\right)e^{2\bar H}\left(e^{4\bar H}-1\right)\nonumber\\
N_{b,1}&=&\left|2+\hat\zeta_1\right|^2\left(\frac{1}{3}+|\hat\zeta_1|^2\right)\left(e^{8\bar H}-1\right)-4\bar H|\hat\zeta_1|^2\left(\frac{5}{3}+2(\hat\zeta_1+\hat\zeta_1^*)+|\hat\zeta_1|^2\right)e^{2\bar H}\left(e^{4\bar H}+1\right)\nonumber\\
&&+8\bar H\left[-\hat\zeta_1(1+\hat\zeta_1)\left(2+2\hat\zeta_1^*+(1+2\bar H^2)(\hat\zeta_1^*)^2\right)+\right.\nonumber\\
&&\qquad \qquad \left.+\left(\frac{1}{3}-\hat\zeta_1^*\right)\left(2+(1-2\bar H^2)|\hat\zeta_1|^2+2(\hat\zeta_1+\hat\zeta_1^*)\right)\right]e^{4\bar H}\nonumber\\
&&+2\left[(1+\hat\zeta_1)(2+\hat\zeta_1)\left(2+2\hat\zeta_1^*+(1+2\bar H^2)(\hat\zeta_1^*)^2\right)+\right.\nonumber\\
&&\qquad \qquad \left.+\left(\frac{1}{3}-\hat\zeta_1^*\right)\left(-2+|\hat\zeta_1|^2+2\bar H^2(2(\hat\zeta_1+\hat\zeta_1^*)+3|\hat\zeta_1|^2)\right)\right]e^{2\bar H}\left(e^{4\bar H}-1\right)\nonumber\\
D_{a,1}&=&(1+\hat\zeta_1)\left[(2+\hat\zeta_1)\left(e^{4\bar H}-1\right)-4\bar H\hat\zeta_1\,e^{2\bar H}\right]\nonumber\\
D_{b,1}&=&\hat\zeta_1(2+\hat\zeta_1)\left(e^{4\bar H}+1\right)+2\left(2+2\hat\zeta_1+(1+2\bar H^2)\hat\zeta_1^2\right)e^{2\bar H} \, ,
\end{eqnarray}
with $\hat\zeta_1=\zeta_1/\eta_1+1/3$ and $\bar H=qH$. We also have, as in the main text,
\begin{equation}
\zeta_1=\frac{\zeta}{1+i\omega\tau_i}\, , \quad \eta_1=\frac{\eta}{1+i\omega\tau_a}\, ,\quad \vartheta_1=\frac{\vartheta}{1+(\omega\tau_i)^2}\, , \quad \theta_1=\frac{\theta}{1+(\omega\tau_a)^2} \, .
\end{equation}
Note that even though it is not directly visible under this form, the expression of $N_{b,1}$ above is real. The linear response function reads
\begin{equation}\label{respComplex}
\chi(q,\omega)=-\frac{1}{q}\left(\frac{1}{\gamma q +\kappa q^3+2i\omega\eta_1 \frac{D_{b,1}}{D_{a,1}}}\right) \, .
\end{equation}
We then have for the effective temperature:
\begin{equation}\label{effTempComplex}
k_BT_{\rm eff}(q,\omega)=\frac{\vartheta_1N_{a,1}+\theta_1N_{b,1}}{2{\rm Re}(\eta_1D_{a,1}^*D_{b,1})} \, .
\end{equation}
We recover the thermodynamic limit of an equilibrium Maxwell model by choosing noise amplitudes that respect detailed balance, $\vartheta=2\zeta k_BT$ and $\theta=2\eta k_BT$. We then have $k_BT_{\rm eff}(q,\omega)\equiv k_BT$, independent of wave vector and frequency.

In general, the effective temperature given by Eq.~(\ref{effTempComplex}) above is a function of frequency and wave vector, as well as of the parameters defining the problem. In the low- and large-$q$ limits, it has two asymptotic finite limits, which have simple expressions. The asymptotic expression in the large-wave-vector regime ($qH\gg 1$) is given in Eq.~(8) of the main text. In the small-wave-vector regime ($qH\ll 1$), we have
\begin{equation}
k_BT_{\rm eff}\simeq\frac{1}{2}\frac{\vartheta_1+\frac{4}{3}\theta_1}{{\rm Re}(\zeta_1+\frac{4}{3}\eta_1)}\, .
\end{equation}
In the low-frequency domain ($\omega\tau_a,\omega\tau_i\ll 1$), the effective temperature is just a function of the wave vector $q$. Its asymptotic expressions read
\begin{equation}
k_BT_{\rm eff}\simeq\frac{\vartheta\eta^2+\theta\left[\left(\zeta+\frac{1}{3}\eta\right)^2+\frac{1}{3}\eta^2\right]}{2\eta\left(\zeta+\frac{1}{3}\eta\right)\left(\zeta+\frac{4}{3}\eta\right)}
\end{equation}
in the large-$q$ limit, and
\begin{equation}\label{eq:effTempLargeQ}
k_BT_{\rm eff}\simeq\frac{1}{2}\frac{\vartheta+\frac{4}{3}\theta}{\zeta+\frac{4}{3}\eta}
\end{equation}
in the low-$q$ limit, as given in Eq.~(12) of the main text.

\section{III. Expressions of the physical quantities for the tissue of finite thickness $H$ in the limit of long timescales}

In the limit of timescales much larger than $\tau_i$ and $\tau_a$, where $\omega\tau_i\ll 1$ and $\omega\tau_a\ll 1$ in the frequency domain, $\zeta_1$ and $\eta_1$ are real and reduce to the bulk and shear viscosities $\zeta$ and $\eta$, respectively. Therefore, the expressions given in section 2 above can be simplified by replacing all the variables by their real, simpler counterparts with $\omega\tau_i\ll 1$ and $\omega\tau_a\ll 1$. We then obtain
\begin{equation}\label{corrReal}
C(q,\omega)=\frac{2}{q}\frac{\vartheta N_a+\theta N_b}{\left| (\gamma q +\kappa q^3)D_a+2i\omega\eta D_b\right|^2} \, ,
\end{equation}
where
\begin{eqnarray}\label{corrQuantitesReal}
N_a&=&\left(2+\hat\zeta\right)^2\left(e^{8\bar H}-1\right)+4\bar H\hat\zeta^2e^{2\bar H}\left(e^{4\bar H}+1\right)+8\bar H\left[2+4\hat\zeta+(1-2\bar H^2)\hat\zeta^2\right]e^{4\bar H}\nonumber\\
&&+2\left(-2+\hat\zeta^2+2\bar H^2\hat\zeta(4+3\hat\zeta)\right)e^{2\bar H}\left(e^{4\bar H}-1\right)\nonumber\\
N_b&=&\left(2+\hat\zeta\right)^2\left(\frac{1}{3}+\hat\zeta^2\right)\left(e^{8\bar H}-1\right)-4\bar H\hat\zeta^2\left(\frac{5}{3}+4\hat\zeta+\hat\zeta^2\right)e^{2\bar H}\left(e^{4\bar H}+1\right)\nonumber\\
&&-\frac{8}{3}\bar H\left[-2+8\hat\zeta+(23+2\bar H^2)\hat\zeta^2+12\hat\zeta^3+(3+6\bar H^2)\hat\zeta^4\right]e^{4\bar H}\nonumber\\
&&+\frac{2}{3}\left[10+4(9+2\bar H^2)\hat\zeta+(31-6\bar H^2)\hat\zeta^2+12\hat\zeta^3+(3+6\bar H^2)\hat\zeta^4\right]e^{2\bar H}\left(e^{4\bar H}-1\right)\nonumber\\
D_a&=&(1+\hat\zeta)\left[(2+\hat\zeta)\left(e^{4\bar H}-1\right)-4\bar H\hat\zeta\,e^{2\bar H}\right]\nonumber\\
D_b&=&\hat\zeta(2+\hat\zeta)\left(e^{4\bar H}+1\right)+2\left(2+2\hat\zeta+(1+2\bar H^2)\hat\zeta^2\right)e^{2\bar H} \, ,
\end{eqnarray}
with $\hat\zeta=\zeta/\eta+1/3$ and $\bar H=qH$. The linear response function reads
\begin{equation}\label{respReal}
\chi(q,\omega)=-\frac{1}{q}\left(\frac{1}{\gamma q +\kappa q^3+2i\omega\eta \frac{D_b}{D_a}}\right) \, .
\end{equation}
We then have for the effective temperature:
\begin{equation}\label{effTempReal}
k_BT_{\rm eff}(q,\omega)=\frac{\vartheta N_a+\theta N_b}{2(\eta D_aD_b)} \, .
\end{equation}
We recover the constant thermodynamic temperature $T$ of a fluid at equilibrium by choosing $\vartheta=2\zeta k_BT$ and $\theta=2\eta k_BT$.

\section{IV. Effective fluctuation-response relation in the limit of long timescales}

\subsection{A. Tissue of finite thickness $H$}

The evolution equation for $\delta H(q,t)$ presented in the main text in the low-frequency domain ($\omega\tau_a,\omega\tau_i\ll 1$) reads
\begin{equation}
\partial_t \delta H(q,t)=-\tau(q)^{-1}\delta H(q,t)+f(q,t)+\xi_{\delta H}(q,t) \, .
\end{equation}
It contains the following functions:
\begin{equation}
\tau(q)=\frac{2\eta}{\gamma q+\kappa q^3}\frac{D_b}{D_a} \qquad \textrm{and} \qquad \xi_{\delta H}(q,t)=\frac{1}{2\eta q}\frac{N_{\delta H}(q,t)}{D_b} \, ,
\end{equation}
where $D_a$ and $D_b$ have already been introduced above in Eq.~(\ref{corrQuantitesReal}). We also have introduced
\begin{eqnarray}
N_{\delta H}(q,t)&=&(2+\hat\zeta)\left[(-2+\hat\zeta)I_1-2J_1\right]e^{4\bar H}+2\left[(2-2\bar H\hat\zeta)I_2+(1+2\bar H)\hat\zeta J_2\right]e^{3\bar H}\nonumber\\
&&+2\left[(2+\hat\zeta+2\bar H\hat\zeta+2(-1+\bar H)\bar H\hat\zeta^2)I_1+(-1+2\bar H)\hat\zeta J_1\right]e^{2\bar H}\nonumber\\
&&+\left[-4(1+\bar H\hat\zeta))I_2+2(2+\hat\zeta) J_2\right]e^{\bar H}-\hat\zeta(2+\hat\zeta)I_1 \, ,
\end{eqnarray}
with
\begin{eqnarray}\label{IIntegrals}
I_1(q,t)&=&-\frac{q}{2}\int_{-H}^{0}dz\,e^{qz}\left(\xi_{xx}-\xi_{zz}+2i\xi_{xy}\right)|_{z}\nonumber\\
I_2(q,t)&=&-\frac{qe^{-\bar H}}{2}\int_{-H}^{0}dz\,e^{-qz}\left(\xi_{xx}-\xi_{zz}-2i\xi_{xy}\right)|_{z}
\end{eqnarray}
and
\begin{eqnarray}\label{JIntegrals}
J_1(q,t)&=&q\int_{-H}^{0}dz\left\{\frac{\hat\zeta q}{2}\int_{-H}^{z}dz'\,e^{qz'}\left(\xi_{xx}-\xi_{zz}+2i\xi_{xy}\right)|_{z'}\right.\nonumber\\
&&\left.-\frac{\hat\zeta q}{2}\int_{z}^{0}dz'\,e^{q(2z-z')}\left(\xi_{xx}-\xi_{zz}-2i\xi_{xy}\right)|_{z'}+e^{qz}\left[-\xi_{zz}+\left(1-\hat\zeta\right)i\xi_{xy}\right]|_{z}\right\}\nonumber\\
J_2(q,t)&=&q e^{-\bar H}\int_{-H}^{0}dz\left\{\frac{\hat\zeta q}{2}\int_{-H}^{z}dz'\,e^{q(-2z+z')}\left(\xi_{xx}-\xi_{zz}+2i\xi_{xy}\right)|_{z'}\right.\nonumber\\
&&\left.-\frac{\hat\zeta q}{2}\int_{z}^{0}dz'\,e^{-qz'}\left(\xi_{xx}-\xi_{zz}-2i\xi_{xy}\right)|_{z'}+e^{-qz}\left[\xi_{zz}+\left(1-\hat\zeta\right)i\xi_{xy}\right]|_{z}\right\} \, ,
\end{eqnarray}
where the noise components are understood as functions of $(q,z,t)$. Dependencies in the coordinate $z$ are indicated as lowerscripts. The equal-time autocorrelation function $\Sigma(q)$, defined such that $\langle \delta H(q,t)\delta H(q',t)\rangle=\Sigma(q)\delta(q+q')$, is then given by
\begin{equation}
\Sigma(q)=\frac{\vartheta N_a+\theta N_b}{2\eta(\gamma q^2 +\kappa q^4)D_aD_b} \, ,
\end{equation}
where $N_a$, $N_b$, $D_a$ and $D_b$ are the functions of $q$ introduced in Eq.~(\ref{corrQuantitesReal}). Finally, the observable $X(q,t)$ defined as
\begin{equation}
X(q,t)=-\tau(q)\Sigma(q)^{-1}\,\delta H(q,t)
\end{equation}
reads
\begin{equation}
X(q,t)=-\frac{4\eta^2q}{\vartheta N_a+\theta N_b}D_b^2\,\delta H(q,t)\, .
\end{equation}
It has the following correlation and response functions in the time domain:
\begin{eqnarray}
C_{XX}(q,t)&=&\frac{8\eta^3}{(\gamma+\kappa q^2)(\vartheta N_a+\theta N_b)}\frac{D_b^3}{D_a}\,e^{-|t|/\tau(q)}\nonumber\\
\chi_{XX}(q,t)&=&-\frac{4\eta^2 q}{\vartheta N_a+\theta N_b}D_b^2\,H(t)e^{-t/\tau(q)} \, ,
\end{eqnarray}
where $H(t)$ is the Heaviside step function. In the frequency domain, they read
\begin{eqnarray}\label{Eq:CorrRespTissueqomega}
C_{XX}(q,\omega)&=&\frac{32\eta^4q}{\vartheta N_a+\theta N_b}\,\frac{D_b^4}{(\gamma q+\kappa q^3)^2D_a^2+4\omega^2\eta^2D_b^2}\nonumber\\
\chi_{XX}(q,\omega)&=&-\frac{8\eta^3q}{\vartheta N_a+\theta N_b}\,\frac{D_b^3}{(\gamma q+\kappa q^3)D_a+2i\omega\eta D_b} \, .
\end{eqnarray}

\subsection{B. Tissue of infinite thickness $H$}

In the limit $qH\gg 1$, the expressions of the quantities defined above can be given explicitly. They read
\begin{equation}
\tau(q)=\frac{2\eta}{\gamma q+\kappa q^3}\frac{\zeta+\frac{1}{3}\eta}{\zeta+\frac{4}{3}\eta} \qquad \textrm{and} \qquad \xi_{\delta H}(q,t)=\frac{1}{2\eta q}\left[\left(1-\frac{2}{\hat\zeta}\right)I_1-\frac{2}{\hat\zeta}J_1\right] \, ,
\end{equation}
with $\hat\zeta=\zeta/\eta+1/3$, and where $I_1$ and $J_1$ are the integrals given by Eqs.~(\ref{IIntegrals}) and (\ref{JIntegrals}) in the limit $qH\gg 1$. This leads to the following expression:
\begin{eqnarray}
\xi_{\delta H}(q,t)&=&-\frac{q}{2\eta}\int_{-\infty}^{0}dz\int_{-\infty}^{z}dz'\,e^{qz'}\left(\xi_{xx}-\xi_{zz}+2i\xi_{xy}\right)|_{z'}\nonumber\\
&&+\frac{q}{2\eta}\int_{-\infty}^{0}dz\int_{z}^{0}dz'\,e^{q(2z-z')}\left(\xi_{xx}-\xi_{zz}-2i\xi_{xy}\right)|_{z'}\nonumber\\
&&+\frac{1}{2\eta}\int_{-\infty}^{0}dz\,e^{qz}\left[\frac{2-\hat\zeta}{2\hat\zeta}\xi_{xx}+\frac{\hat\zeta+2}{2\hat\zeta}\xi_{zz}+i\xi_{xy}\right]|_{z}\, .
\end{eqnarray}
We then have
\begin{equation}
\Sigma(q)=\frac{1}{\gamma q^2+\kappa q^4}\frac{\vartheta\eta^2+\theta\left[\left(\zeta+\frac{1}{3}\eta\right)^2+\frac{1}{3}\eta^2\right]}{2\eta\left(\zeta+\frac{1}{3}\eta\right)\left(\zeta+\frac{4}{3}\eta\right)}
\end{equation}
and
\begin{equation}
X(q,t)=-\frac{4\eta^2(\zeta+\frac{1}{3}\eta)^2}{\vartheta\eta^2+\theta\left[(\zeta+\frac{1}{3}\eta)^2+\frac{1}{3}\eta^2\right]}q\delta H(q,t)\, ,
\end{equation}
which gives the following correlation and response functions in the time domain:
\begin{eqnarray}
C_{XX}(q,t)&=&\frac{8}{\gamma+\kappa q^2}\frac{\left(\zeta+\frac{1}{3}\eta\right)^2}{\vartheta\eta^2+\theta\left[\left(\zeta+\frac{1}{3}\eta\right)^2+\frac{1}{3}\eta^2\right]}\frac{\zeta+\frac{1}{3}\eta}{\zeta+\frac{4}{3}\eta}\,e^{-|t|/\tau(q)}\nonumber\\
\chi_{XX}(q,t)&=&-4q\frac{\left(\zeta+\frac{1}{3}\eta\right)^2}{\vartheta\eta^2+\theta\left[\left(\zeta+\frac{1}{3}\eta\right)^2+\frac{1}{3}\eta^2\right]}\,H(t)e^{-t/\tau(q)}\, ,
\end{eqnarray}
and
\begin{eqnarray}\label{Eq:CorrRespTissueqomega}
C_{XX}(q,\omega)&=&\frac{32\eta^4q}{\vartheta\eta^2+\theta\left[\left(\zeta+\frac{1}{3}\eta\right)^2+\frac{1}{3}\eta^2\right]}\frac{\left(\zeta+\frac{1}{3}\eta\right)^4}{\left(\zeta+\frac{4}{3}\eta\right)^2}\frac{1}{\left[(\gamma q+\kappa q^3)^2+4\omega^2\tilde\eta^2\right]}\nonumber\\
\chi_{XX}(q,\omega)&=&-\frac{8\eta^3q}{\vartheta\eta^2+\theta\left[\left(\zeta+\frac{1}{3}\eta\right)^2+\frac{1}{3}\eta^2\right]}\frac{\left(\zeta+\frac{1}{3}\eta\right)^3}{\zeta+\frac{4}{3}\eta}\frac{1}{\gamma q+\kappa q^3+2i\omega\tilde\eta}
\end{eqnarray}
in the frequency domain. As in the main text, $\tilde\eta=\eta(\zeta+\eta/3)/(\zeta+4\eta/3)$.

\section{V. Estimation of the collision length and time}

We now estimate the collision length $l_c$ defined in the main text in the low-frequency, small-wave-vector domain by either Eq.~(14) or Eq.~(15) of the main text, depending on the value of $\gamma H^2/(k_BT_{\rm eff})$. In the low-frequency, small-wave-vector domain, $k_BT_{\rm eff}$ is given by Eq.~(12) of the main text, and we start by estimating this quantity. To estimate the tissue bulk viscosity $\zeta$, we use experiments where the bulk growth rate $k_p=k_d-k_a$ of a tissue spheroid is determined as a function of the externally applied stress thanks to a fitting procedure~\cite{Montel:2011uq,montel2012isotropic}. The measurement procedure leads to $\zeta=-\left(\partial k_p/\partial P\right)^{-1}\simeq 5\,10^4$~Pa$\cdot$day. With a cell-division rate $k_d$ of the order of one per day, we get an estimate for the isotropic elastic modulus of the spheroid in its homeostatic state $\bar{E}=\zeta k_d$ of the order of $5\,10^4$~Pa. This is in agreement with another study, in which the short-time responses of multicellular spheroids to an external pressure jump are measured, and which give $\bar{E}\simeq 10^4-10^5$~Pa~\cite{delarue2014stress}. The anisotropic counterparts of these quantities are the tissue shear viscosity $\eta$, the shear elastic modulus $E$ and the rate of cell-neighbor exchange $k_r$, related by $E\simeq\eta k_r$. Estimating that $E$ is typically lower if not much lower than $\bar{E}$~\cite{Forgacs:1998bh,Guevorkian:2010ye} and that $k_r$ should typically be much larger than $k_d$, we get that the tissue bulk viscosity $\zeta$ should be much larger than the tissue shear viscosity $\eta$. This is in agreement with experimental estimates of the shear viscosity of cellular aggregates, which give $\eta\simeq 10^4-10^5$~Pa$\cdot$s~\cite{Forgacs:1998bh,Beysens:2000zr,Marmottant:2009lh,Guevorkian:2010ye}. The amplitude $\vartheta$ of the isotropic noise due to cell-division and apoptosis is given in the main text. In the homeostatic state where $k_d=k_a$, it reads $\vartheta=2\zeta^2k_d/\rho_h=2\bar{E}^2/(k_d\rho_h)$. Estimating a similar expression for the anisotropic noise amplitude $\theta\simeq\eta^2k_r/\rho_h\simeq E^2/(k_r\rho_h)$, we see that $k_d\ll k_r$ together with $\bar{E}\gtrsim E$ or $\bar{E}\gg E$ imposes $\vartheta\gg\theta$. With these estimates, we therefore have
\begin{equation}
k_BT_{\rm eff}\simeq\frac{1}{2}\frac{\vartheta}{\zeta}=\frac{\zeta k_d}{\rho_h}\, .
\end{equation}

We now turn to the proper estimation of the collision length $l_c$. We introduce the length $w=\gamma/(\zeta k_d)$ and the linear cell dimension $a$ with $\rho_h=a^{-3}$. With a typical tissue thickness $H=10\,a$, we have $l_c\simeq l_\kappa \exp(100\pi w/a)$ in the tension-dominated regime and $l_c\simeq 10\,l_\kappa \sqrt{(w/a)}$ in the bending-dominated regime. Within this context, the ratio $l_c/l_\kappa$ is therefore function of the adimensional parameter $w/a$. A reasonable scale for $a$ is 10~$\mu$m. Tissue surface tensions have been measured for different tissue types and using different experimental methods. The measurements give values ranging typically from a fraction up to several Millinewton per meter~\cite{fung1993biomechanics,Foty:1994dq,Foty:1996cr,Forgacs:1998bh,Beysens:2000zr,Foty:2005fk,Schotz:2008qo,Mgharbel:2009qf,Guevorkian:2010ye,Gonzalez-Rodriguez:2012fk}. With these values and $\zeta k_d\simeq 10^4-10^5$~Pa as estimated above, one finds that $w/a$ ranges typically between $10^{-4}$ and $10^{-2}$, at most $10^{-1}$. The lower bound of this interval corresponds to the bending-dominated regime, which leads to $l_c\simeq 0.1\,l_\kappa$. The upper bound corresponds to the tension-dominated regime, with $l_c$ ranging from a few tens of $l_\kappa$ to being much larger than $l_\kappa$. These order-of-magnitude estimates therefore indicate that actual tissues could be in either of these two regimes, depending on parameter values. In addition, in the tension-dominated regime, the collision length $l_c$ is exponentially dependent on the parameters characterizing the tissue. We therefore expect that small variations in some of the parameters entering the expression of $l_c$, like the tissue surface tension $\gamma$, can have dramatic effects on the structure of the tissue surface, and that this could happen independently of alterations of the cell-division rate.

To complete our estimates, we can finally give an order of magnitude for the tissue effective temperature that we have defined. Using the values given above, one finds $k_BT_{\rm eff}\simeq 10^{-11}-10^{-10}$~J, a value that is some 10 orders of magnitude larger than thermal energy. This should however not surprise us, since the stochastic processes we study here are related to the behavior of whole cells, which are typically five orders of magnitude larger than individual molecules.

\end{large}

\end{widetext}

\end{document}